\documentclass[aps,pra,twocolumn,groupedaddress,superscriptaddress]{revtex4-2}
\usepackage{graphicx}
\usepackage{amssymb,amsmath}
\usepackage{xcolor}
\usepackage{bm}
\usepackage{hyperref}
\usepackage{ulem}
\usepackage{siunitx}
\newcommand{\ml}{m_{\ell}}

\begin{document}

\title{Orbital-Dependent Dimensional Crossover of a $p$-Wave Feshbach Resonance}
%\title{Orbital-Selective Control of a $p$-Wave Feshbach Resonance via Dimensional Crossover in a 1D Optical Lattice}

\author{Hang Yu}
\thanks{These authors contributed equally to this work.}
\affiliation{School of Physics and Astronomy, Sun Yat-Sen University, Zhuhai 519082, China}
\author{Liao Sun}
\thanks{These authors contributed equally to this work.}
\affiliation{School of Physics and Astronomy, Sun Yat-Sen University, Zhuhai 519082, China}
\author{Shaokun Liu}
\affiliation{School of Physics and Astronomy, Sun Yat-Sen University, Zhuhai 519082, China}
\author{Shuai Peng}
\affiliation{School of Physics and Astronomy, Sun Yat-Sen University, Zhuhai 519082, China}

\author{Jiaming Li}
%\thanks{These authors contributed equally to this work.}
\email[]{lijiam29@mail.sysu.edu.cn}
\affiliation{School of Physics and Astronomy, Sun Yat-Sen University, Zhuhai 519082, China}
\affiliation{Guangdong Provincial Key Laboratory of Quantum Metrology and Sensing, Sun Yat-Sen University, Zhuhai 519082, China}
\affiliation{Shenzhen Research Institute of Sun Yat-Sen University, Shenzhen 518057, China}
\affiliation{State Key Laboratory of Optoelectronic Materials and Technologies, Sun Yat-Sen University, Guangzhou 510275, China}

\author{Le Luo}
\email[]{luole5@mail.sysu.edu.cn}
\affiliation{School of Physics and Astronomy, Sun Yat-Sen University, Zhuhai 519082, China}
\affiliation{Guangdong Provincial Key Laboratory of Quantum Metrology and Sensing, Sun Yat-Sen University, Zhuhai 519082, China}
\affiliation{Shenzhen Research Institute of Sun Yat-Sen University, Shenzhen 518057, China}
\affiliation{State Key Laboratory of Optoelectronic Materials and Technologies, Sun Yat-Sen University, Guangzhou 510275, China}

\date{\today}
%--------------------------------------------------------------------------%

\begin{abstract}

We report the observation of a dimensional crossover of a $p$-wave Feshbach resonance in an ultracold, spin-polarized $^6$Li Fermi gas confined by a one-dimensional optical lattice. 
Using high-resolution atom-loss spectroscopy, we resolve the orbital doublet associated with the $\ml=0$ and $|\ml|=1$ scattering channels over a wide range of lattice depths.
In the weak-confinement regime, the atom loss signal associated with the $|\ml|=1$ branch is stronger, consistent with the twofold orbital degeneracy of the three-dimensional system. 
As the lattice confinement increases, the relative loss weight of the two orbital branches evolves continuously toward the quasi-two-dimensional limit, indicating a progressive suppression of relative motion along the lattice direction.
In addition, we observe a systematic confinement dependence of the orbital splitting between the two resonance branches. 
These results provide an experimental characterization of orbital-dependent $p$-wave scattering in reduced dimensions and motivate future microscopic studies of confined anisotropic scattering.

\end{abstract}

\maketitle

%%%%%%%%%%%%%%%%%%%%%%%%%%%%%%%%%%%%%%%%%%%%%%%%%%%%
\section{Introduction}

\begin{figure*}[htbp]
\begin{center}
	\includegraphics[width=2\columnwidth, angle=0, scale=1.0]{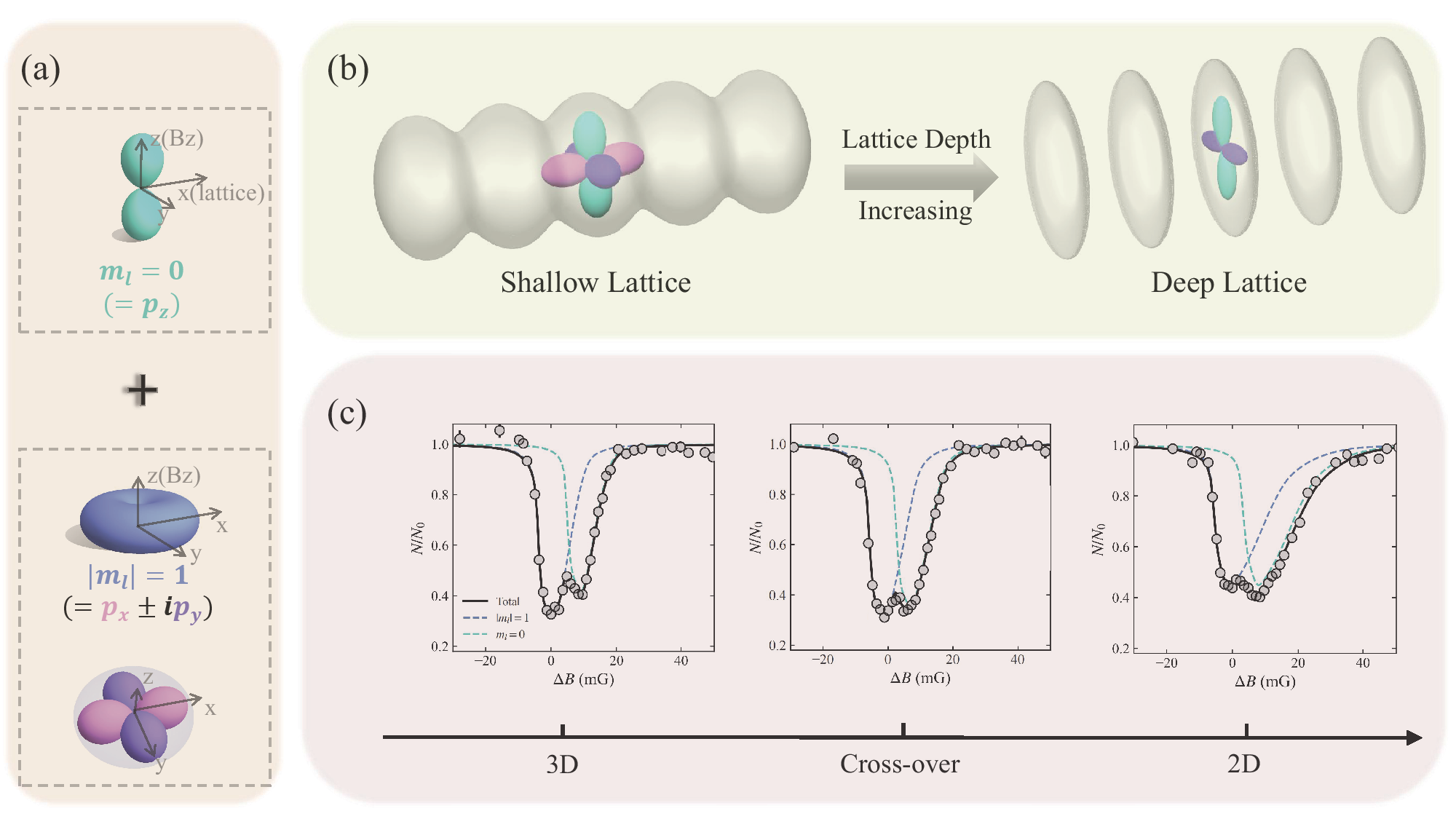}
	\vspace{-5pt}
	\caption{
		Schematic of dimensional crossover and orbital anisotropy of $p$-wave interactions in a one-dimensional optical lattice.
		(a) Spatial orientation of the $p$-wave scattering orbitals. The magnetic field defines the quantization axis ($z$), while the lattice wave vector is along $x$. The $\ml=0$ component corresponds to the $p_z$ orbital (teal) aligned with the $z$ axis. The $|\ml|=1$ branch is represented by a blue doughnut-shaped distribution, while the bottom panel illustrates its decomposition into the two Cartesian components, $p_x$ (pink) and $p_y$ (purple).
		(b) Geometry of the lattice confinement. Increasing the lattice depth drives the system from a weakly modulated 3D geometry (shallow lattice) to an array of isolated quasi-2D pancake-shaped sites (deep lattice). Strong confinement along the lattice axis suppresses the $p_x$ contribution within the $|\ml|=1$ orbitals, reducing their relative scattering weight compared with the orthogonal $\ml=0$ ($p_z$) component.
		(c) Evolution of the atom-loss spectrum across the dimensional crossover. Representative spectra are shown close to the quasi-3D limit (left, $R = 1.94 \pm 0.25$), crossover region (middle, $R = 1.39 \pm 0.14$), and quasi-2D limit (right, $R = 1.08 \pm 0.10$). 
		Gray circles represent the experimental data, with error bars (less than 4\%) smaller than the circle size. Solid black curves are total fits, and colored dashed lines indicate contributions from individual orbital branches ($\ml=0$ in teal, and $|\ml|=1$ in blue).
	}    
	\label{p:Schematic Diagram}
\end{center}
\end{figure*}

Unlike $s$-wave scattering, $p$-wave interactions access collision channels with nonzero orbital angular momentum and intrinsic anisotropy~\cite{ticknor2004multiplet,Ho2005,Gurarie2005}.
This orbital degree of freedom enables rich few-body and many-body physics, including orbital-selective interactions and topological pairing states~\cite{gurarie2007,Sato2017,Buhler2014}.
Near a magnetic Feshbach resonance, the anisotropy of $p$-wave scattering is amplified, producing a dependence on the relative orientation between the colliding particles and the quantization axis even at ultralow energies~\cite{Chin2010,Peng2025,Nagase2026}.
Consequently, multiple scattering channels distinguished by the magnetic quantum number $\ml$ naturally emerge.

The energy splitting between different $\ml$ channels provides spectroscopic access to orbital-resolved scattering processes and manifests as a characteristic double-peak structure in macroscopic observables, such as atom loss induced by inelastic collisions~\cite{Suno2003,huckans2009threebody,regal2003tuning,gerken2019,Shu2024,zhang2004p}.
By analyzing both the relative positions and amplitudes of these loss features, one can access information about the orbital structure of a $p$-wave Feshbach resonance and its coupling to the scattering continuum.

A central question is how this orbital-resolved structure can be controlled in a systematic and tunable manner.
Beyond magnetic-field tuning, dimensional confinement provides a geometry-based way to reshape the scattering environment and modify orbital selectivity~\cite{Olshanii1998,gunter2005p,Venu2023,Chang2020,Waseem2016,Naidon2017,Zhou2019,Zhou2021,Jackson2023,Dale2024}.
In an optical lattice, confinement induces anisotropy in the relative kinetic energy, which couples directly to the orbital symmetry of $p$-wave scattering.
As illustrated in Fig.~\ref{p:Schematic Diagram}, increasing lattice depth in a one-dimensional (1D) lattice drives the system from a quasi-three-dimensional (3D) regime with substantial intersite tunneling to a quasi-two-dimensional (2D) geometry of isolated pancake-shaped sites~\cite{Martiyanov2010}, progressively suppressing relative motion along the lattice axis and redistributing the orbital contributions to scattering.

While reduced dimensionality is known to modify $p$-wave scattering~\cite{Pricoupenko2008,Idziaszek2009,Tan2014,Kurlov2017}, establishing a direct connection between these confinement effects and experimentally accessible orbital-resolved observables remains challenging.
This is particularly true for the narrow $p$-wave resonance in $^6$Li~\cite{Note_Gao}, where the intrinsic $m_l$ splitting is only a few milligauss and therefore comparable to the energy scales introduced by confinement.
Although dimensional crossover effects have been explored previously in $^6$Li and other atomic species~\cite{gunter2005p,guo2024}, the confinement dependence of the orbital branching ratio and orbital splitting has not been systematically investigated.

Here we experimentally investigate how dimensional confinement modifies a $p$-wave Feshbach resonance in a spin-polarized $^6$Li Fermi gas confined in a 1D optical lattice near 159~G.
Using high-resolution atom-loss spectroscopy together with an effective analysis framework for the three-body loss spectra, we track the evolution of the resonance doublet associated with the $\ml=0$ and $|\ml|=1$ scattering channels as the system evolves from the quasi-3D regime toward the quasi-2D limit.
By analyzing both the orbital branching ratio and the orbital splitting, we observe a continuous evolution of the orbital-resolved resonance structure, including a redistribution of the orbital loss weight and a systematic increase of the splitting between the $|\ml|=1$ and $\ml=0$ resonance branches as the system crosses over from the 3D toward the quasi-2D limit.
These observations reveal how dimensional confinement modifies the orbital-resolved structure of the resonance.
Our results provide a quantitative experimental characterization of how dimensional confinement reshapes the orbital-resolved structure of narrow $p$-wave Feshbach resonances.

%Because the 1D optical lattice explicitly breaks the 3D rotational symmetry, we describe the scattering channels in the Cartesian basis ($p_x, p_y, p_z$) instead of the conventional $\ml$ basis. Specifically, the $\ml=0$ channel corresponds to the $p_z$ orbital along the magnetic field, while the $|\ml|=1$ manifold comprises the initially degenerate $p_x$ and $p_y$ orbitals (hereafter denoted as $p_{x,y}$).

\section{Experiment}
\label{sec:Exp}

Following the experimental procedure described in Ref.~\cite{Liu2026}, we prepare a 3D ultracold, two-component Fermi gas of $^6$Li atoms in the two-lowest hyperfine state $|1\rangle \equiv |F=1/2, m_F=1/2\rangle$ and $|2\rangle \equiv |F=1/2, m_F=-1/2\rangle$ at a magnetic field of 320~G.
The gas has a temperature of $T/T_\mathrm{F}\approx 0.5$, where $T_\mathrm{F}$ is the Fermi temperature and a typical atom number of $N \approx 3\times10^5$.

The atoms are subsequently loaded into a 1D optical lattice oriented along the $x$ axis.
The lattice is formed by two laser beams with a wavelength of $\lambda = 1064~\mathrm{nm}$ intersecting at an angle of $2\theta = 20^{\circ}$.
This configuration produces a lattice with a spacing of
$d = \lambda / (2\sin\theta)=3.1~\mathrm{\mu m}$, together with an anisotropic trapping potential characterized by trap-frequency ratios $\omega_x : \omega_z : \omega_y \approx 514 : 5.8 : 1$.
The lattice depth $V_0$, parameterized by the dimensionless depth $s = V_0/E_R$ with a recoil energy of $E_R/h = 883.5~\mathrm{Hz}$, is continuously tunable up to $s = 60$ by adjusting the laser power using an acousto-optic modulator.
Further details of the 1D lattice setup and lattice loading procedure can be found in Refs.~\cite{Gong2023}.

A homogeneous magnetic field $B$ is applied along the $z$ axis, perpendicular to the lattice direction.
In this configuration, the $\ml=0$ scattering channel is mainly associated with the orbital component aligned with the magnetic field (denoted here as $p_z$, teal), while the $|\ml|=1$ channels correspond to transverse orbital motion in the $x$–$y$ plane and are formed by the combinations $p_x \pm i p_y$ (blue), as illustrated in Fig.~\ref{p:Schematic Diagram}(a).

Atom-loss spectroscopy is performed near the $p$-wave Feshbach resonance in the $|1\rangle$–$|1\rangle$ scattering channel, located at $B_0 \approx 159.227~\mathrm{G}$~\cite{Note_Liu}. 
After loading the atoms into the lattice at a given depth $V_0$, the magnetic field is first ramped to $164~\mathrm{G}$ and held for $100~\mathrm{ms}$ to allow rethermalization. 
To bypass eddy-current-induced magnetic-field distortions and accurately control the interaction time, we implement a radio-frequency (RF) gated method~\cite{Liu2026}. 
Specifically, a $50~\mu\mathrm{s}$ resonant light pulse is applied to clear the $|1\rangle$ atoms, leaving a spin-polarized $|2\rangle$ gas that is immune to the $|1\rangle$–$|1\rangle$ resonance. 
The magnetic field is then swept to a target value $B = B_0 + \Delta B$, where $\Delta B$ denotes the magnetic-field detuning. 
The atoms are kept in state $|2\rangle$ to allow the induced eddy currents to fully decay.
Once the magnetic field stabilizes, a $2.5~\mathrm{ms}$ RF pulse transfers the atoms into the $|1\rangle$ state, initiating the interaction duration $t_{\mathrm{hold}} = 350~\mathrm{ms}$. 
Throughout this interaction window, the magnetic field is actively stabilized using a low-noise coil system with current feedback~\cite{Chen2021, Liu2023} together with $50\text{-}\mathrm{Hz}$ noise compensation~\cite{Peng2025}. 
This system enables a minimum field step of $1.3~\mathrm{mG}$ and achieves an rms field fluctuation of $0.1~\mathrm{mG}$, providing sufficient resolution to track the $\sim 8~\mathrm{mG}$ splitting between the $|\ml|=1$ and $\ml=0$ resonances.

During the $350~\mathrm{ms}$ interaction window, three-body recombination leads to observable atom loss. This holding duration $t_{\mathrm{hold}}$ is optimized to maximize the signal-to-noise ratio of the loss spectra; typically, about $30\%\text{--}40\%$ of the atoms remain at the resonance peaks, providing sufficient spectral contrast to resolve the orbital splitting.
After the three-body reaction at the target field, the remaining atom number $N(t_{\mathrm{hold}})$ is measured by absorption imaging after time-of-flight expansion following a rapid magnetic-field switch back to 164~G. Details of the temperature determination in this dimensional-crossover study are provided in Appendix~\ref{sm:temperature}. 

%By repeating this sequence for different magnetic fields, we reconstruct the atom-loss spectrum $N(B)$; Fig.~\ref{p:Schematic Diagram}(c) shows three representative measured spectra, in which two distinct $p$-wave loss features associated with the $p_z$ (right peak) and $p_{x,y}$ (left peak) channels are clearly resolved~\cite{Liu2026}.
%Furthermore, by tuning the lattice depth $V_0$ from shallow ($s \approx 12$) to deep ($s \approx 60$), we systematically investigate how the relative contributions of the two orbital branches evolve as the system crosses over from weak to strong confinement.

\section{Data analysis}
\label{sec:DataAnalysis}

We develop a cascade three-body model to evaluate the atom-loss spectrum, where two $|1\rangle$ atoms first form a Feshbach dimer through two-body $p$-wave resonance scattering, followed by atom-dimer relaxation that characterizes the inelastic atom loss. 

We start from the low-energy quasi-2D $p$-wave scattering amplitude~\cite{Idziaszek2009,Kurlov2017},
\begin{equation}
f^{2\mathrm{D}}_p(q)
=
\frac{4 q^2}
{
	A_p^{-1}
	+ B_p q^2
	- \frac{2}{\pi} q^2 \ln (l_x q)
	+ i q^2
},
\label{eq:f2Dp}
\end{equation}
where $q$ is the relative momentum and
$l_x = \sqrt{\hbar/(m\omega_x)}$ is the harmonic oscillator length associated with the lattice confinement.
The effective 2D scattering area $A_p$ and effective range parameter $B_p$ are determined by the confinement geometry according to
$A_p^{-1} = (4 / 3\sqrt{2\pi} l_x^2)( l_x^3/V_p + k_e l_x/2 - c_1 )$
and
$B_p = (4/3\sqrt{2\pi})( l_x k_e - c_2 )$,
where $k_e$ is the 3D $p$-wave effective range parameter,
and $c_1 \approx 0.0655$ and $c_2 \approx 0.1464$.

The confinement-shifted resonance position $B_{\mathrm{res}}$ is determined by the pole condition $A_p^{-1}=0$.
Parameterizing the scattering volume near resonance as
$V_p(B) \approx -V_{\mathrm{bg}} \Delta / (B - B_0)$,
we obtain the confinement-induced shift~\cite{Waseem2016}
\begin{equation}
B_{\mathrm{res}}
=
B_0
-
\frac{c_1 V_{\mathrm{bg}} \Delta}{l_x^3}
+
\frac{k_e V_{\mathrm{bg}} \Delta}{2 l_x^2}.
\label{eq:shift_model}
\end{equation}
Here we use the known $^6$Li parameters
$V_{\mathrm{bg}} \approx -4.236\times10^4\,a_0^3$,
$\Delta \approx -40.13~\mathrm{G}$,
and $k_e = 0.151 \pm 0.006\,a_0^{-1}$~\cite{Peng2025}.
Here, $B_{\mathrm{res}}$ denotes the resonance position predicted by the effective two-body scattering model, while the experimentally extracted loss maxima are used as spectroscopic observables to characterize their evolution under confinement.

The confinement-induced shift in Eq.~(\ref{eq:shift_model})  follows the standard confinement-induced-resonance (CIR) picture.
In the quasi-2D geometry, the relative motion along the confined direction acquires an additional zero-point energy, which shifts the scattering threshold upward and consequently modifies the magnetic-field position at which the molecular state becomes resonant with the scattering continuum.
This approximation is applicable to the present narrow, closed-channel-dominated $p$-wave resonance in $^6$Li. As the confinement strength increases, the relative collision threshold is shifted upward, leading to a corresponding shift of the resonance position in magnetic field. Within this effective description, the confinement shift is therefore expected to be approximately common to different orbital channels at leading order.

Then, following Ref.~\cite{Ngampruetikorn2013}, the three-body loss coefficient $K_3(E)$ is related to the atom-dimer inelastic scattering cross-section. To effectively describe the inelastic relaxation, we construct the atom-dimer scattering amplitude by introducing a phenomenological imaginary part to the inverse scattering area $1/A_p^{\mathrm{ad}} \to 1/A_p^{\mathrm{ad}} + i/A_i^{\mathrm{ad}}$, inspired by the two-body treatment in Ref.~\cite{Kurlov2017}. 

A microscopic implementation of this approach would require the atom-dimer scattering parameters, which are not available for the present system. Motivated by the observation of Ref.~\cite{Ngampruetikorn2013} that the energy dependence of three-body loss follows that of the underlying two-body resonance, we approximate the atom-dimer resonance profile using the corresponding two-body parameters ($A_p$, $A_i$, $E_b$, and $\Gamma_{\mathrm{in}}$), determined by the experimentally known quantities $V_{\mathrm{bg}}$, $\Delta B$, and $k_e$.

Within this effective description, the resonance position and spectral line shape are inherited from the underlying two-body scattering properties. Since the short-range atom-dimer relaxation process is not treated microscopically, the overall magnitude of the loss is introduced through a phenomenological scaling factor $\kappa$. The resulting effective expression is
\begin{equation}
\begin{split}
	K_3(E) =
	\frac{(36\pi \hbar^3/m^2 E) \times 2 \kappa \Gamma_e(E) \Gamma_{\mathrm{in}}}
	{
		\left[E - E_b - \Gamma_e(E) C(E)\right]^2
		+
	 \Gamma^2_{tot}(E)
	},
\end{split}
\label{eq:K3_final}
\end{equation}
where $C(E)=\frac{1}{\pi}\ln(l_x q)$, the binding energy $E_b=-\hbar^2/(m A_p B_p)$ and the inelastic width $\Gamma_{\mathrm{in}}=2\hbar^2/(m B_p A_i)$ characterize the atom-dimer resonance, yielding a total width $\Gamma_{\mathrm{tot}}(E)=\Gamma_e(E)+\Gamma_{\mathrm{in}}$.
Details on the interpretation of $A_i$ and $\kappa$, as well as the robustness analysis of the fitting, are provided in Appendix~\ref{sm:kappa} and \ref{sm:correlation}, respectively.

To connect the energy-dependent loss coefficient to the observed atom-loss dynamics, we thermally average $K_3(E)$ over a Boltzmann distribution, following the procedure of Ref.~\cite{Liu2026}. 
This yields the effective three-body loss coefficients $Q_3^0$ and $Q_3^1$ for the $m_l=0$ and $|m_l|=1$ channels, respectively. 
The resulting atom-loss dynamics within each lattice site are then described by the three-body rate equation,
\begin{equation}
\frac{dN_s}{dt}
=
-\frac{1}{3(2\pi\sigma_r^2)^2}
\Big[
r_1 Q_3^{1} + (1-r_1) Q_3^{0}
\Big]
N^3_s.
\label{eq:rate_equation}
\end{equation}
Here, $N_s$ denotes the atom number per individual lattice site, and the geometric factor $(2\pi\sigma_r^2)^2$ originates from integrating the thermal Gaussian density profile with an effective radial width $\sigma_r$, and $r_1$ denotes the fractional contribution of the $|\ml|=1$ channel. 

%In the analysis of the experimental data, Eq.~(\ref{eq:All}) is solved numerically to account for magnetic-field noise and other experimental imperfections.
%For the idealized case of a constant loss coefficient, however, 
Equation~(\ref{eq:rate_equation}) admits an analytical solution,
\begin{equation}
N_s(t)
=
\frac{N_s(0)}
{\sqrt{
		1 +
		\dfrac{2 t}{3(2\pi\sigma_r^2)^2}
		\Big[
		r_1 Q_3^{1} + (1-r_1) Q_3^{0}
		\Big]
		N_s(0)^2
}},
\label{eq:fit}
\end{equation}
where $N_s(0)$ is the initial atom number. Detailed derivations of these governing equations are provided in Appendix~\ref{sm:model}.

\begin{figure}[!htbp]
\begin{center}
	\includegraphics[width=\columnwidth, angle=0, scale=1.0]{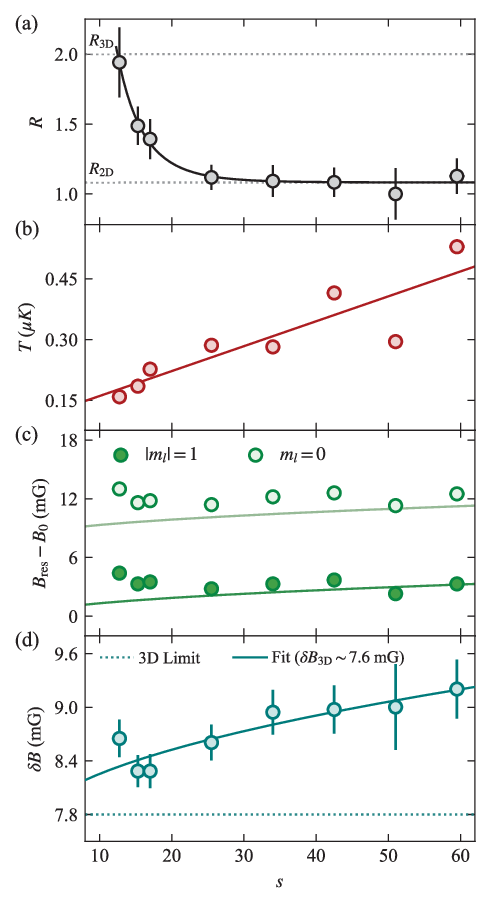}
	\vspace{-15pt}
	\caption{
		Branching ratio $R$ (a), the measured temperature $T_\mathrm{tof}$ (b), resonance positions (c), and resonance splitting $\delta B$ (d) versus lattice depth $s$, with solid lines indicating fits or model-based calculations as described in the text. The deviation at $s<20$ likely originates from the breakdown of the simple 2D confinement model in the 2D-3D crossover; the splitting between resonances, however, remains accurate due to cancellation of common-mode uncertainties. The horizontal dotted line in (d) indicates the experimentally measured 3D limit $\delta B_{\mathrm{3D}} = 7.6~\mathrm{mG}$~\cite{Liu2026}.
	}    \label{p:shift}
\end{center}
\end{figure}	

\section{Results}
\label{sec:result}

We measure the magnetic-field-dependent atom-loss spectra at different lattice depths $s$ using a fixed holding time $t_{\mathrm{hold}}=350$~ms. 
The spectra are fitted using Eq.~(\ref{eq:fit}), with the relative loss weight $r_1$ treated as a free parameter. 
Representative fitting results are shown in Fig.~\ref{p:shift}.

We define the orbital branching ratio as $R \equiv r_1 / (1-r_1)$, and summarize the extracted values in Fig.~\ref{p:shift}(a). 
In shallow lattices, a typical atom-loss spectrum is presented on the left side of Fig.~\ref{p:Schematic Diagram}(c), where the $|\ml|=1$ component (left peak) exhibits a stronger loss signal than the $\ml=0$ component (right peak)\cite{Peng2025}. 
In this weak-confinement regime, scattering motion along the lattice direction remains active, so that both the $p_x$ and $p_y$ orbitals contribute to the $|\ml|=1$ channel, as illustrated on the left side of Fig.~\ref{p:Schematic Diagram}(b), while the $\ml=0$ channel originates solely from the $p_z$ orbital. Assuming comparable scattering strength among the three orbital components $p_x$, $p_y$, and $p_z$, the twofold degeneracy of the $|\ml|=1$ channel leads to approximately twice the loss strength of the $\ml=0$ channel. Consequently, the branching ratio approaches the isotropic 3D limit $R \rightarrow 2$~\cite{Liu2026}.

In deep lattices, as presented on the right side of Fig.~\ref{p:Schematic Diagram}(c), the loss strengths of the $|\ml|=1$ and $\ml=0$ components become comparable. 
Here, scattering motion along the lattice direction is suppressed, which reduces the contribution of the $p_x$ orbital from the $|\ml|=1$ component, as illustrated on the right side of Fig.~\ref{p:Schematic Diagram}(b). 
As a result, the remaining $p_y$ contribution produces a loss strength comparable to that of the $p_z$ orbit, driving the branching ratio toward the quasi-2D limit $R \rightarrow 1$. 
Note that the unequal peak heights in the spectra originate from asymmetric thermal broadening on the BCS side of the resonance~\cite{Peng2024cp}.

Between these two limits, $R$ evolves continuously from 2 toward 1 as $s$ increases. A representative intermediate-depth spectrum is shown in the middle of Fig.~\ref{p:Schematic Diagram}(c). 
This behavior reflects the progressive suppression of tunneling-assisted motion along the lattice direction. 
To characterize this crossover behavior, we fit the data using the phenomenological tight-binding form
\begin{equation}
R(s) = R_{\mathrm{2D}} + A_\mathrm{R} \exp\!\left[-\gamma \sqrt{s}\right],
\end{equation}
where $A_\mathrm{R}$ is a free amplitude parameter.
The fit yields an asymptotic quasi-2D limit $R_{\mathrm{2D}} = 1.08 \pm 0.02$ and a decay constant $\gamma = 2.01 \pm 0.27$. The extracted $\gamma$ is consistent with the theoretical scaling exponent of the lowest-band tunneling amplitude, $J(s) \propto \exp(-2\sqrt{s})$~\cite{Zwerger2003}.

Moreover, from the measured temperature $T_{\mathrm{tof}}$ shown in Fig.~\ref{p:shift}(b), we find that although $T_{\mathrm{tof}}$ gradually increases with lattice depth, the ratio $k_B T_{\mathrm{tof}}/\hbar\omega_x$ remains below unity throughout the explored range, indicating the suppression of axial motion. 
Together, the observed evolution of $R$ and the relevant thermodynamic energy scales suggest that the dimensional crossover cannot be understood solely in terms of local single-site confinement. 
Instead, the results are more naturally interpreted as a consequence of the gradual suppression of intersite tunneling with increasing lattice depth. 
The relevant energy scales and dimensional-crossover criteria are discussed in Appendix~\ref{sm:scale}, while the comparison with these two phenomenological models is presented in Appendices~\ref{sm:single_site} and~\ref{sm:tunneling}.

Besides modifying $R$, the confinement also shifts the resonance positions and modifies the orbital splitting. 
The extracted resonance positions $B_{\mathrm{res}}$, referenced to the 3D resonance position $B_0$, are shown in Fig.~\ref{p:shift}(c). 
The solid curves correspond to the confinement-induced threshold shifts predicted by Eq.~(\ref{eq:shift_model}).

Although Eq.~(\ref{eq:shift_model}) is formally derived in the quasi-2D limit, it provides a useful effective description of the dominant confinement-induced resonance shift throughout the dimensional crossover.
This is consistent with the standard confinement-induced-resonance picture, in which transverse confinement primarily modifies the open-channel scattering threshold through the additional zero-point energy. The agreement improves with increasing lattice depth, while deviations become apparent in the shallow-lattice regime ($s<20$).

These deviations likely reflect the crossover nature of the system, where axial motion is not yet fully frozen and inter-site tunneling remains significant. In this regime, effects beyond the quasi-2D description, including coupling between center-of-mass and relative motion that can give rise to inelastic confinement-induced resonances (ICIRs)~\cite{Haller2010,Saenz2012,Saenz2016}, may contribute to the observed behavior.

For the narrow, closed-channel-dominated ${}^6\mathrm{Li}$ $p$-wave resonance studied here, the dominant confinement effect is expected to arise from the modification of the scattering threshold. While the observed loss spectra are well described by the resonance-doublet structure, we emphasize that our measurements primarily characterize the evolution of orbital-resolved loss spectra under confinement. 
The precise relationship between these observed loss features and the underlying two-body scattering poles remains an open question requiring future microscopic theory.

The extracted orbital splitting $\delta B$ is shown in Fig.~\ref{p:shift}(d). 
Similarly to $B_\mathrm{res}$, the orbital splitting $\delta B$ exhibits a weak but systematic increase with lattice depth. 
Since the splitting is extracted directly from individual double-peak spectra, common-mode uncertainties, such as magnetic-field noise and calibration offsets, largely cancel, allowing the relative peak separation to be determined with higher precision than $B_\mathrm{res}$.

To characterize this evolution, we also fit the splitting using the phenomenological form
\begin{equation}
\delta B(s)=\delta B_{\mathrm{3D}}+A_B\sqrt{s}.
\end{equation}
The fit gives $\delta B_{\mathrm{3D}}=7.6\pm0.3~\mathrm{mG}$ and $A_B=0.2\pm0.1 ~ \mathrm{mG}$, where $\delta B_{\mathrm{3D}}$ is consistent with the previous measurement in a 3D gas~\cite{Liu2026}. 

\begin{figure}[!htbp]
\begin{center}
	\includegraphics[width=\columnwidth, angle=0, scale=1.0]{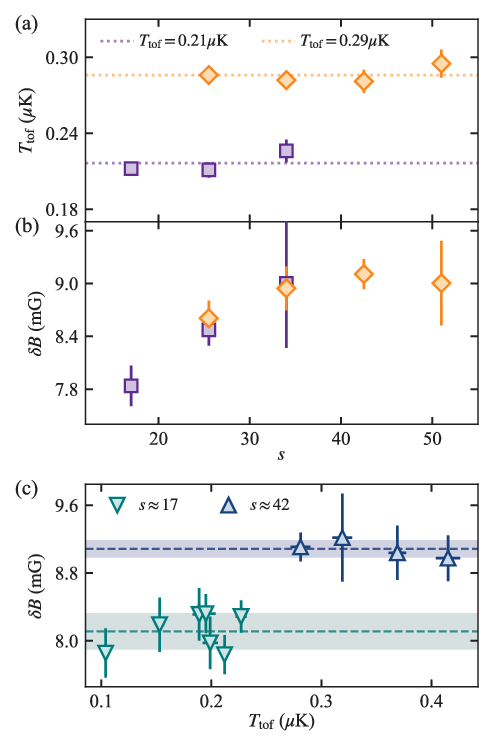}
	\vspace{-15pt}
	\caption{
		Extra results for the dependence of resonance splitting $\delta B$. 
		(a) $T_{\mathrm{tof}}$ versus $s$ to verify that the temperature remains constant. 
		(b) $\delta B$ versus $s$ at these temperatures.
		(c) $\delta B$ versus $T_{\mathrm{tof}}$ at fixed lattice depths.
		The dashed lines indicate mean values of $8.07\,\mathrm{mG}$ for $s \approx 17$ and $9.17\,\mathrm{mG}$ for $s \approx 42$.
		Error bars and shaded regions represent the standard deviation.
	}    \label{p:comp_T}
\end{center}
\end{figure}

To further investigate the confinement dependence of $\delta B$, we performed additional measurements summarized in Fig.~\ref{p:comp_T}. 
First, by varying the initial depth of the 3D optical dipole trap, we prepared samples with comparable temperatures at different lattice depths as shown in Fig.~\ref{p:comp_T}(a). 
Owing to the reduced evaporative cooling efficiency and additional lattice-induced heating at large $s$, the achievable temperature stabilization is limited to a relatively narrow parameter range.

Figure~\ref{p:comp_T}(b) compares the extracted splitting $\delta B$ for two representative temperatures, $T\approx0.21~\mu$K and $0.29~\mu$K. 
For the higher temperature, $\delta B$ gradually saturates at approximately $9~\mathrm{mG}$ for $s\gtrsim30$. 
This behavior suggests that finite-temperature broadening and the associated collision-energy distribution begin to limit the further evolution of the spectral features in the deep-lattice regime. 
In contrast, at the lower temperature, $\delta B$ continues to evolve more noticeably over the range $15\lesssim s\lesssim35$. 
These observations indicate that the primary evolution of the splitting is driven by lattice confinement, while finite-temperature effects mainly influence its saturation behavior at large lattice depths.

Second, we directly varied the temperature at fixed lattice depths $s=17$ and $42$, as shown in Fig.~\ref{p:comp_T}(c). 
Within experimental uncertainty, the extracted splitting $\delta B$ remains unchanged over the explored temperature range for both lattice depths. 
This result further supports that the observed enhancement of the orbital splitting is predominantly associated with confinement rather than thermal effects.

The observed evolution of $\delta B$ suggests that the splitting between the $|m_l|=1$ and $m_l=0$ branches cannot be fully understood as a simple common-mode confinement shift of the resonance threshold. 
At the same time, the relatively small magnitude of the effect and the crossover nature of the present system prevent a quantitative microscopic interpretation within the effective description used here. 
A more complete understanding will likely require a multichannel treatment of confined $p$-wave scattering beyond the quasi-2D approximation.

\section{Conclusion and Discussion}

Using high-stability atom-loss spectroscopy in a 1D optical lattice, we resolve the orbital doublet of the $^6$Li $p$-wave Feshbach resonance and follow its evolution across the dimensional crossover from quasi-3D to quasi-2D.
The two resonance branches remain spectrally distinguishable over a wide range of lattice depths, enabling a systematic study of both their relative loss strengths and their confinement-dependent resonance positions.

We observe that the orbital branching ratio between the $|\ml|=1$ and $\ml=0$ channels evolves continuously with increasing lattice depth, approaching the isotropic 3D limit under weak confinement and the quasi-2D limit under strong confinement. 
This evolution is consistent with the progressive suppression of scattering motion along the lattice direction and is consistent with a confinement-induced redistribution of orbital scattering weight.

In addition, we observe a systematic evolution of the resonance splitting $\delta B$ with increasing lattice depth. 
Temperature-dependent measurements indicate that this behavior is primarily associated with lattice confinement.
At the same time, the relatively small magnitude of the effect and the crossover nature of the present system prevent a definitive microscopic interpretation within the current framework.

Our results provide an experimental characterization of orbital-dependent $p$-wave scattering in reduced dimensions and may motivate future microscopic studies of confined anisotropic interactions and low-dimensional many-body phenomena.

% \section{Code availability}
% \textit{Code availability.} Relevant code for data analysis are available in the text. Additional software used in this study is available from the corresponding authors upon reasonable request.

\section{Acknowledgements}
This work is supported by NSFC under Grant  No. 12174458 and No. 12574302. J. Li received support from the Fundamental Research Funds for the Central Universities, Sun Yat-sen University (24xkjc015). L. Luo received support from Shenzhen Science and Technology Program
JCYJ20220818102003006.

\section{Data availability}
The original data are provided in the Appendix. Any additional data can be requested by e-mailing the corresponding author.

%\section{Author contributions}

% \section{Competing interests}
% \textit{Competing interests.} The authors declare that they have no conflict of interest

% \bibliography{reference}

\begin{thebibliography}{99}

\bibitem{ticknor2004multiplet}
C. Ticknor, C. A. Regal, D. S. Jin, and J. L. Bohn, ``Multiplet structure of Feshbach resonances in nonzero partial waves'', \textit{Phys. Rev. A} \textbf{69}, 042712 (2004).

\bibitem{Ho2005}
T.-L. Ho and R. B. Diener, ``Fermion Superfluids of Nonzero Orbital Angular Momentum near Resonance'', \textit{Phys. Rev. Lett.} \textbf{94}, 090402 (2005).

\bibitem{Gurarie2005}
V. Gurarie, L. Radzihovsky, and A. V. Andreev, ``Quantum phase transitions across a p-wave Feshbach resonance'', \textit{Phys. Rev. Lett} \textbf{94}, 230403 (2005).

\bibitem{gurarie2007}
V. Gurarie and L. Radzihovsky, ``Resonantly paired fermionic superfluids'', \textit{Ann. Phys.} \textbf{322}, 2--119 (2007).

\bibitem{Sato2017}
M. Sato and Y. Ando, ``Topological superconductors: a review'', \textit{Rep. Prog. Phys.} \textbf{80}, 76501 (2017).

\bibitem{Buhler2014}
A. B{\"u}hler, N. Lang, C. V Kraus, G. M{\"o}ller, S. D Huber, and H.-P. B{\"u}chler, ``Majorana modes and p-wave superfluids for fermionic atoms in optical lattices'', \textit{Nat. Commun.} \textbf{5}, 4504 (2014).

\bibitem{Chin2010}
C. Chin, R. Grimm, P. Julienne, and E. Tiesinga, ``Feshbach resonances in ultracold gases'', \textit{Rev. Mod. Phys.} \textbf{82}, 1225--1286 (2010).

\bibitem{Peng2025}
S. Peng, S. Peng, L. Ren, S. Liu, B. Liu, J. Li, and L. Luo, ``Precision Measurement of Spin-Dependent Dipolar Splitting in $^{6}\mathrm{Li}$ $p$-Wave Feshbach Resonances'', \textit{Phys. Rev. Lett.} \textbf{135}, 133401 (2025).

\bibitem{Nagase2026}
K. Nagase, H. Takahashi, S. Oshima, and T. Mukaiyama, ``Temperature Dependence of $p$-Wave Contacts in a Harmonically Trapped Fermi Gas'', \textit{Phys. Rev. Lett.} \textbf{136}, 013402 (2026).

\bibitem{Suno2003}
H. Suno, B. D. Esry, and C. H. Greene, ``Recombination of Three Ultracold Fermionic Atoms'', \textit{Phys. Rev. Lett.} \textbf{90}, 053202 (2003).

\bibitem{huckans2009threebody}
J. H. Huckans, J. R. Williams, E. L. Hazlett, R. W. Stites, and K. M. O'Hara, ``Three-body recombination in a three-state Fermi gas with widely tunable interactions'', \textit{Phys. Rev. Lett} \textbf{102}, 165302 (2009).

\bibitem{regal2003tuning}
C. A. Regal, C. Ticknor, J. L. Bohn, and D. S. Jin, ``Tuning $p$-Wave Interactions in an Ultracold Fermi Gas of Atoms'', \textit{Phys. Rev. Lett.} \textbf{90}, 053201 (2003).

\bibitem{gerken2019}
M. Gerken, B. Tran, S. H\"afner, E. Tiemann, B. Zhu, and M. Weidem\"uller, ``Observation of dipolar splittings in high-resolution atom-loss spectroscopy of $^{6}\mathrm{Li}$ $p$-wave Feshbach resonances'', \textit{Phys. Rev. A} \textbf{100}, 050701 (2019).

\bibitem{Shu2024}
S. Peng, T. Shu, B. Si, S. Peng, Y. Guo, Y. Han, J. Li, G. Wang, and L. Luo, ``Observation of a broad state-to-state spin-exchange collision near a $p$-wave Feshbach resonance of $^{6}\mathrm{Li}$ atoms'', \textit{Phys. Rev. A} \textbf{110}, L051301 (2024).

\bibitem{zhang2004p}
J. Zhang, E. G. M. van Kempen, T. Bourdel, L. Khaykovich, J. Cubizolles, F. Chevy, M. Teichmann, L. Tarruell, S. J. J. M. F. Kokkelmans, and C. Salomon, ``$P$-wave Feshbach resonances of ultracold $^{6}\mathrm{Li}$'', \textit{Phys. Rev. A} \textbf{70}, 030702 (2004).

\bibitem{Olshanii1998}
M. Olshanii, ``Atomic Scattering in the Presence of an External Confinement and a Gas of Impenetrable Bosons'', \textit{Phys. Rev. Lett.} \textbf{81}, 938--941 (1998).

\bibitem{gunter2005p}
K. G\"unter, T. St\"oferle, H. Moritz, M. K\"ohl, and T. Esslinger, ``$p$-Wave Interactions in Low-Dimensional Fermionic Gases'', \textit{Phys. Rev. Lett.} \textbf{95}, 230401 (2005).

\bibitem{Venu2023}
V. Venu, P. Xu, M. Mamaev, F. Corapi, T. Bilitewski, J. P. D'Incao, C. J. Fujiwara, A. M. Rey, and J. H. Thywissen, ``Unitary p-wave interactions between fermions in an optical lattice'', \textit{Nature} \textbf{613}, 262--267 (2023).

\bibitem{Chang2020}
Y.-T. Chang, R. Senaratne, D. {Cavazos-Cavazos}, and R. G. Hulet, ``Collisional Loss of One-Dimensional Fermions Near a p -Wave Feshbach Resonance'', \textit{Phys. Rev. Lett.} \textbf{125}, 263402 (2020).

\bibitem{Waseem2016}
M. Waseem, Z. Zhang, J. Yoshida, K. Hattori, T. Saito, and T. Mukaiyama, ``Creation of p-wave Feshbach molecules in selected angular momentum states using an optical lattice'', \textit{J. Phys. B} \textbf{49}, 204001 (2016).

\bibitem{Naidon2017}
P. Naidon and S. Endo, ``Efimov physics: a review'', \textit{Rep. Prog. Phys.} \textbf{80}, 056001 (2017).

\bibitem{Zhou2019}
M. He and Q. Zhou, ``$s$-wave contacts of quantum gases in quasi-one-dimensional and quasi-two-dimensional traps'', \textit{Phys. Rev. A} \textbf{100}, 012701 (2019).

\bibitem{Zhou2021}
M. He and Q. Zhou, ``$p$-wave contacts of quantum gases in quasi-one-dimensional and quasi-two-dimensional traps'', \textit{Phys. Rev. A} \textbf{104}, 043303 (2021).

\bibitem{Jackson2023}
K. G. Jackson, C. J. Dale, J. Maki, K. G. S. Xie, B. A. Olsen, D. J. M. Ahmed-Braun, S. Zhang, and J. H. Thywissen, ``Emergent $s$-Wave Interactions between Identical Fermions in Quasi-One-Dimensional Geometries'', \textit{Phys. Rev. X} \textbf{13}, 021013 (2023).

\bibitem{Dale2024}
C. J. Dale, K. G. S. Xie, K. Pond Grehan, S. Zhang, J. Maki, and J. H. Thywissen, ``Emergent $s$-wave interactions in orbitally active quasi-two-dimensional Fermi gases'', \textit{Phys. Rev. A} \textbf{110}, L051302 (2024).

\bibitem{Martiyanov2010}
K. Martiyanov, V. Makhalov, and A. Turlapov, ``Observation of a Two-Dimensional Fermi Gas of Atoms'', \textit{Phys. Rev. Lett.} \textbf{105}, 030404 (2010).

\bibitem{Pricoupenko2008}
L. Pricoupenko, ``Resonant Scattering of Ultracold Atoms in Low Dimensions'', \textit{Phys. Rev. Lett.} \textbf{100}, 170404 (2008).

\bibitem{Idziaszek2009}
Z. Idziaszek, ``Analytical solutions for two atoms in a harmonic trap: $p$-wave interactions'', \textit{Phys. Rev. A} \textbf{79}, 062701 (2009).

\bibitem{Tan2014}
S.-G. Peng, S. Tan, and K. Jiang, ``Manipulation of $p$-Wave Scattering of Cold Atoms in Low Dimensions Using the Magnetic Field Vector'', \textit{Phys. Rev. Lett.} \textbf{112}, 250401 (2014).

\bibitem{Kurlov2017}
D. V. Kurlov and G. V. Shlyapnikov, ``Two-body relaxation of spin-polarized fermions in reduced dimensionalities near a $p$-wave Feshbach resonance'', \textit{Phys. Rev. A} \textbf{95}, 032710 (2017).

\bibitem{Gao2011}
B. Gao, ``Analytic description of atomic interaction at ultracold temperatures. II. Scattering around a magnetic Feshbach resonance'', \textit{Phys. Rev. A} \textbf{84}, 022706 (2011).

\bibitem{Note_Gao}
We note that the resonance is classified as ``narrow'' based on its small resonance-strength parameter ($\xi_{\mathrm{res}} \approx 0.19$) rather than its magnetic-field linewidth~\cite{Gao2011}. 

\bibitem{guo2024}
Y. Guo, H. Yao, S. Ramanjanappa, S. Dhar, M. Horvath, L. Pizzino, T. Giamarchi, M. Landini, and H.-C. N{\"a}gerl, ``Observation of the 2D--1D crossover in strongly interacting ultracold bosons'', \textit{Nat. Phys.} \textbf{20}, 934--938 (2024).

\bibitem{Liu2026}
S. Liu, Z. Xu, S. Peng, S. Peng, T. Shu, J. Li, and L. Luo, ``Orbital-resolved three-body recombination across a p-wave Feshbach resonance in ultracold 6Li'', \textit{Rep. Prog. Phys.} \textbf{89}, 020502 (2026).

\bibitem{Gong2023}
H. Gong, H. Liu, B. Jiao, H. Zhang, H. Yu, Q. Peng, S. Peng, T. Shu, Y. Zhu, J. Li, and L. Luo, ``Controllable production of degenerate Fermi gases of $^{6}\mathrm{Li}$ atoms in the crossover from two dimensions to three dimensions'', \textit{Phys. Rev. A} \textbf{107}, 053321 (2023).

\bibitem{Note_Liu}
The slight discrepancy between this value and the resonance position previously reported by our group in Ref.~\cite{Liu2026} is primarily attributed to variations in the temperature of the magnetic field coils.

\bibitem{Chen2021}
Y. Chen, S. Peng, H. Gong, X. Zhang, J. Li, and L. Luo, ``Characterization of the magnetic field through the three-body loss near a narrow Feshbach resonance'', \textit{Phys. Rev. A} \textbf{103}, 063311 (2021).

\bibitem{Liu2023}
H. Liu, S. Peng, B. Jiao, J. Li, and L. Luo, ``Ultra-low noise bipolar current source for ultracold atom magnetic system'', \textit{Rev. Sci. Instrum.} \textbf{94}, 053201 (2023).

\bibitem{Ngampruetikorn2013}
V. Ngampruetikorn, M. M. Parish, and J. Levinsen, ``Three-body problem in a two-dimensional Fermi gas'', \textit{Europhys. Lett.} \textbf{102}, 13001 (2013).

\bibitem{Peng2024cp}
S. Peng, H. Liu, J. Li, and L. Luo, ``Collisional cooling of a Fermi gas with three-body recombination'', \textit{Commun. Phys.} \textbf{7}, 101 (2024).

\bibitem{Zwerger2003}
W. Zwerger, ``Mott--Hubbard transition of cold atoms in optical lattices'', \textit{J. Opt. B: Quantum Semiclass. Opt} \textbf{5}, S9 (2003).

\bibitem{Haller2010}
E. Haller, M. J. Mark, R. Hart, J. G. Danzl, L. Reichs\"ollner, V. Melezhik, P. Schmelcher, and H.-C. N\"agerl, ``Confinement-Induced Resonances in Low-Dimensional Quantum Systems'', \textit{Phys. Rev. Lett.} \textbf{104}, 153203 (2010).

\bibitem{Saenz2012}
S. Sala, P.-I. Schneider, and A. Saenz, ``Inelastic Confinement-Induced Resonances in Low-Dimensional Quantum Systems'', \textit{Phys. Rev. Lett.} \textbf{109}, 073201 (2012).

\bibitem{Saenz2016}
S. Sala and A. Saenz, ``Theory of inelastic confinement-induced resonances due to the coupling of center-of-mass and relative motion'', \textit{Phys. Rev. A} \textbf{94}, 022713 (2016).

\end{thebibliography}

\newpage

\clearpage

\appendix

\iffalse
\setcounter{equation}{0} 
\setcounter{figure}{0}   
\setcounter{table}{0}    

\renewcommand{\theequation}{A\arabic{equation}} 
\renewcommand{\thefigure}{S\arabic{figure}}  
\renewcommand{\thetable}{A\arabic{table}}    

\makeatletter
\renewcommand{\theHequation}{A\arabic{equation}}
\renewcommand{\theHfigure}{S\arabic{figure}}
\renewcommand{\theHtable}{A\arabic{table}}
\makeatother
\fi

\section{Time-of-Flight Temperature Measurement}
\label{sm:temperature}

In the presence of axial lattice confinement, the atomic motion along the lattice direction is modified. 
We therefore characterize the thermodynamic state of the gas using temperatures extracted from time-of-flight expansion measurement in the radial axis, which primarily probe the in-plane kinetic-energy distribution.

The temperature $T_{\mathrm{tof}}$ is obtained by measuring the transverse cloud size $\sigma_z(\tau)$ after different expansion times $\tau$ and fitting the ballistic expansion using
\begin{equation}
\sigma_z^2(\tau) = \sigma_{z0}^2 + \frac{k_B T_{\mathrm{tof}}}{m}\tau^2,
\end{equation}
where $\sigma_{z0}$ is the in-situ transverse size and $m$ is the atomic mass.

The extracted temperature $T_{\mathrm{tof}}$ serves as an operational thermodynamic scale throughout this work. 
In particular, it is used to evaluate the ratio
$\hbar\omega_x/k_B T_{\mathrm{tof}}$,
which characterizes the relative strength of axial confinement.

\begin{figure}[htbp]
\begin{center}
	%\vspace{-0.3cm}
	\includegraphics[width=\columnwidth, angle=0, scale=1.0]{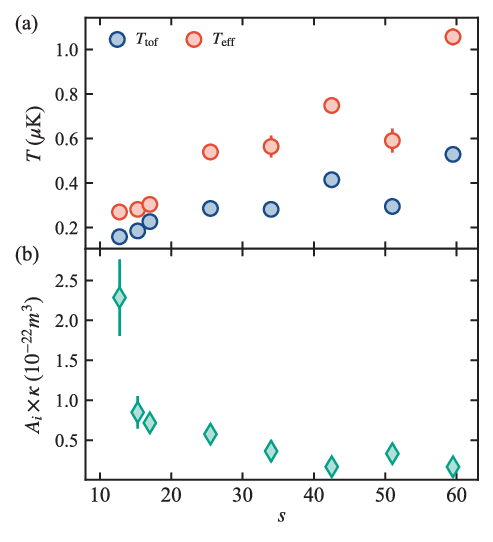}
	\vspace{-15pt}
	\caption{
		(a) Comparison of the measured time-of-flight temperature $T_{\mathrm{tof}}$ (blue circles) and the effective fit temperature $T_{\mathrm{eff}}$ (red circles) as a function of lattice depth $s$.
		While there is a systematic offset, both temperatures exhibit a consistent increasing trend with lattice depth. (b) The fitted product $A_i \times \kappa$ shows a monotonic decrease with increasing lattice depth $s$, indicating a reduced effective loss strength in the deep-lattice regime.
	}    
	\label{p:Teff}
\end{center}
\end{figure}

As shown in Fig.~\ref{p:Teff}(a), $T_{\mathrm{tof}}$ increases with lattice depth. 
This trend originates from finite trap-depth effects and lattice-induced heating, which preferentially retain higher-energy atoms at larger lattice depths.

In addition to $T_{\mathrm{tof}}$, we introduce an effective temperature $T_{\mathrm{eff}}$ extracted from fits to the atom-loss spectra. 
The resulting $T_{\mathrm{eff}}$ values, also shown in Fig.~\ref{p:Teff}(a), exhibit a proportional dependence on $T_{\mathrm{tof}}$. 
This consistency indicates that the thermal broadening extracted from the loss spectra follows the same trend as the independently measured kinetic-energy scale.

\section{More Details of the Effective Three-Body Loss Model}
\label{sm:model}

This section provides additional details of the effective three-body loss model used in the main text, including the construction of the loss coefficient, the thermal averaging procedure, and the connection to the macroscopic loss dynamics.

We model the observed atom loss as a cascade three-body process: two colliding $|1\rangle$ atoms first form a resonant $p$-wave Feshbach dimer through two-body scattering, followed by inelastic atom-dimer relaxation that removes particles from the trap. 
A similar mechanism was recently shown to successfully describe three-body loss in the 3D $|1\rangle$ channel of $^6$Li~\cite{Liu2026}.

Following Ref.~\cite{Ngampruetikorn2013}, the energy-dependent three-body loss coefficient $K_3(E)$ is related to the atom-dimer inelastic scattering cross-section $\sigma^{\mathrm{inel}}_{p,\mathrm{ad}}(E)$ through
\begin{equation}
K_3(E)
=
\frac{24\pi\hbar^2}{mE}
v_{\mathrm{ad}}
\sigma^{\mathrm{inel}}_{p,\mathrm{ad}}(E),
\label{eq:K3_SM}
\end{equation}
where $v_\mathrm{ad}$ is the atom-dimer relative velocity. 
The inelastic cross-section is given by $\sigma^{\mathrm{inel}}_{p,\mathrm{ad}} = \frac{2}{q}(1-|S^{2\mathrm{D}}_{p,\mathrm{ad}}|^2)$, with the 2D S-matrix $S_{p,\mathrm{ad}}^{2\mathrm{D}} = 1 + f_{p,\mathrm{ad}}^{2\mathrm{D}}(q)/2i$ and the collision energy $E=\hbar^2 q^2/m$.

Although Eq.~\ref{eq:K3_SM} is proposed in Ref.~\cite{Ngampruetikorn2013}, the explicit format of the inelastic scattering cross-section is lacking. Here, we adopt the method in Ref.~\cite{Kurlov2017} to construct the inelastic atom-dimer scattering amplitude by adding an imaginary contribution to the inverse scattering area,
$1/A_p^{\mathrm{ad}} \to 1/A_p^{\mathrm{ad}} + i/A_i^{\mathrm{ad}}$, where $A_i^{\mathrm{ad}}$ quantifies the relaxation of the dimer into deeply bound states through collisions. 
This leads to the effective Breit--Wigner form:
\begin{equation}
f^{2\mathrm{D}}_{p,\mathrm{ad}}(E)
=
\frac{2 \Gamma_e(E)}
{
	E - E_b^\mathrm{ad} - \Gamma_e(E) C(E)
	+ i \Gamma_{\mathrm{tot}}^{ad}(E)/2
},
\label{eq:f2Dp_res}
\end{equation}
where $C(E)=\frac{1}{\pi}\ln(l_x q)$. Here, the binding energy $E_b^\mathrm{ad}=-\hbar^2/(m A_p^{\mathrm{ad}} B_p)$ and the inelastic width $\Gamma_{\mathrm{in}}^\mathrm{ad}=2\hbar^2/(m B_p A_i^{\mathrm{ad}})$ characterize the atom-dimer resonance, yielding a total width $\Gamma_{\mathrm{tot}}^\mathrm{ad}(E)=\Gamma_e(E)+\Gamma_{\mathrm{in}}^\mathrm{ad}$.

It is important to note that Ref.~\cite{Kurlov2017} focuses on two-body inelastic decay. Here, we extend this treatment to the atom-dimer relaxation process. This requires knowledge of the atom-dimer scattering parameters $A_p^{\mathrm{ad}}$ and $A_i^{\mathrm{ad}}$.
Ref.~\cite{Ngampruetikorn2013} suggests that the energy dependence of the three-body loss follows that of the underlying two-body resonance. Motivated by this observation, we approximate the atom-dimer resonance profile using the corresponding two-body parameters $A_p$, $A_i$, $E_b$, and $\Gamma_{\mathrm{in}}$, which are determined from $V_{\mathrm{bg}}$, $\Delta$, and $k_e$.

Because this substitution is only approximate, it does not reproduce the absolute magnitude of the three-body loss. 
We therefore introduce a scaling factor $\kappa$, giving the effective expression used in Eq.~(\ref{eq:K3_final}) of the main text. 

For the multi-parameter fit, we treat the resonance positions of the two orbital channels as free parameters. They are defined via the resonance position $B_{\mathrm{res}}$ and the orbital splitting $\delta B$ such that $B_{\mathrm{res}}^1 = B_{\mathrm{res}}$ and $B_{\mathrm{res}}^0 = B_{\mathrm{res}} + \delta B$, where the superscript ``1'' labels the $|\ml|=1$ branch and the superscript ``0'' labels the $\ml=0$ branch.  The binding energy $E_b$ in Eq.~(\ref{eq:K3_final}) is then calculated for each channel using its respective resonance position.

The thermally averaged three-body loss coefficient
$Q_3(T,B,B_{\mathrm{res}})$
is obtained by integrating $K_3(E)$ over a Boltzmann energy distribution,
\begin{equation}
Q_3(T,B,B_{\mathrm{res}})
=
\frac{3}{k_B T}
\int_0^\infty
K_3(E)
e^{-E/(k_B T)}
\, dE.
\label{eq:Q3_integral}
\end{equation}
Here the energy-dependent loss rate $K_3(E)$ implicitly depends on the magnetic field $B$ through the detuning from  $B_{\mathrm{res}}$, and this dependence is suppressed in the notation for simplicity.

Although the initial gas is moderately degenerate ($T/T_F\sim0.5$), we employ Boltzmann thermal averaging. This approximation is justified by the centrifugal barrier of $p$-wave scattering, which strongly suppresses low-momentum collisions. As a result, the thermally averaged loss is dominated by the higher-energy part of the distribution, where the difference between Boltzmann and Fermi--Dirac statistics is small~\cite{Peng2025,Liu2026}. In the actual fitting, the temperature entering this thermal average is taken as the effective fit temperature $T_{\mathrm{eff}}$.

The two branch-resolved loss coefficients are denoted as
$Q_3^0 = Q_3(T,B,B_{\mathrm{res}}^0)$ and 
$Q_3^1 = Q_3(T,B,B_{\mathrm{res}}^1)$, 
corresponding to the $\ml=0$ and $|\ml|=1$ channels, respectively.
The dimensionless parameter $r_1$ ($0 \le r_1 \le 1$) characterizes the fractional contribution of the $p_{x,y}$ channel to the total three-body loss. In the fit, $r_1$ is not fixed to the ideal degeneracy value, but is extracted from the measured double-peak spectrum at each lattice depth.

To convert the thermally averaged loss coefficients into the measured remaining atom number, we use the rate equation given in Eq.~(\ref{eq:rate_equation}) of the main text.
The geometric factor $(2\pi\sigma_r^2)^2$ originates from the mean-square atomic
density per lattice site.
In the radial directions, the atomic density within each site is well
approximated by a thermal Gaussian distribution with an effective width
$\sigma_r$, where $\sigma_r^2 =k_B T_{\mathrm{eff}}/(m \omega_r^2)$, $\omega_r = \sqrt{\omega_y \omega_z}$, $m$ is the atomic mass. The mean atom number per lattice site $N_s$ is obtained by dividing the total atom number $N$ by the number of occupied lattice sites, which is 27 in our experiment. The analytical solution, Eq.~(\ref{eq:fit}) of the main text, is then used to fit the remaining atom number after the fixed hold time.

%Finally, we check whether the large atom loss affects the extracted resonance centers. Although a macroscopic atom loss of up to $\sim 70\%$ exceeds the linear-response regime, our extracted orbital splitting $\delta B$ relies solely on the relative separation of resonance centers. Measurements at different holding times verify that while peak depths and linewidths vary with the total loss, the fitted center positions and $\delta B$ remain constant within experimental uncertainty. This check indicates that nonlinear loss dynamics do not produce a resolvable systematic shift in the extracted resonance positions.

\section{Lattice-depth dependence of the fitted loss strength}
\label{sm:kappa}

In the analysis of the atom-loss spectra using Eq.~\ref{eq:fit} in the main text, the product $A_i \times \kappa$ serves as an effective fitting parameter characterizing the overall magnitude of the inelastic loss. 
Since $\kappa$ is introduced phenomenologically to account for the approximate treatment of the atom-dimer resonance, the resonance position and linewidth are determined by the underlying two-body parameters, while the extracted quantity $A_i \times \kappa$ should not be interpreted as a microscopic three-body loss coefficient.

Figure~\ref{p:Teff}(b) shows that the fitted value of $A_i \times \kappa$ decreases monotonically with increasing lattice depth $s$. 
This behavior is consistent with the experimentally observed reduction of atom loss under stronger lattice confinement and suggests a gradual suppression of inelastic collision processes in the deep-lattice regime.

\section{Robustness of Multi-parameter Fitting}
\label{sm:correlation}

In our multi-parameter fitting, the initial atom number $N_s(0)$ is fixed to the independently measured value obtained at a far-detuning magnetic field. The other parameters, $B_{\mathrm{res}}$, $\delta B$, $A_i$, $r_1$, $T_{\mathrm{eff}}$ and $\kappa$, are allowed to vary within broad bounds. For each individual curve, we test different combinations of initial guesses, and all tested initializations lead to the same best-fit solution reported in the main text.

We further extract the covariance matrix in the vicinity of the optimal solution and normalize it to obtain the parameter correlation matrix. The resulting matrices show similar overall patterns across different lattice depths.
For the two fitting parameters discussed in the main text, $R$ and $\delta B$, the absolute values of each correlation-matrix element with the remaining fitting parameters are typically below $0.5$, suggesting that neither of them exhibits strong local correlations with the others within the local linear approximation around the optimum.

We emphasize that, since the loss features are not symmetric, the overlap between the two branches is absorbed in the double-Gaussian fitting mainly through unequal fitted widths of the two components and thus the double-Gaussian fitting is not used to check the robustness. The analysis therefore relies solely on the physical loss model.

\section{Thermodynamic and dynamical scales across the dimensional crossover}
\label{sm:scale}

To quantitatively characterize the confinement regime explored in this work, the relevant thermodynamic energy scales and dynamical timescales are summarized in Table~\ref{tab:parameter_crossover_SM}. 

As the lattice depth increases, the axial confinement energy $\hbar\omega_x$ exceeds both the thermal energy $k_B T_{\mathrm{tof}}$ and the Fermi energy $E_F$ throughout the explored parameter range. 
This indicates that axial excitation is energetically suppressed under lattice confinement. 
However, a thermodynamic criterion alone is insufficient to fully characterize the dimensional crossover, and it is therefore necessary to further consider the dynamical properties of the system through the competition between the inter-site tunneling time $\tau_J = h/4J$ and the inelastic collision timescale $\tau_{\mathrm{coll}}$~\cite{gunter2005p}. 

Using the measured three-body loss, $\tau_{\mathrm{coll}}$ is estimated from the observed survival fraction ($30\%\text{--}40\%$ remaining after a holding time $t_{\mathrm{hold}}=350$~ms), yielding $\tau_{\mathrm{coll}}\approx69\text{--}134$~ms near resonance.

At shallow lattice depths such as $s=12$, the tunneling time remains relatively short ($\tau_J\approx23.5$~ms $< \tau_{\mathrm{coll}}$), allowing substantial interlayer coupling and effectively 3D motion. 
In contrast, at deep lattice depths such as $s=34$, the tunneling time becomes much longer ($\tau_J\approx1.04$~s $\gg \tau_{\mathrm{coll}}$), rendering inter-site tunneling dynamically irrelevant on the timescale of the inelastic collisions. 
Under these conditions, the gas behaves approximately as a set of decoupled quasi-2D layers.

Using the condition $\tau_{\mathrm{coll}}\sim\tau_J$ as an estimate for the crossover between effectively 3D and quasi-2D dynamical behavior, we obtain a characteristic lattice-depth range of $s \approx \mathrm{12}\text{--}\mathrm{34}$. 
This estimate agrees well with the experimentally observed crossover region identified from the evolution of the orbital branching ratio $R$ in Fig.~\ref{p:shift}(a) of the main text.

\begin{table*}[htbp]
\centering
\caption{Summary of experimental parameters for the data points presented in Fig.~\ref{p:shift}. The table lists the lattice depth $s$, the corresponding trapping frequency $\omega_{x}/2\pi$, transverse confinement energy $\hbar\omega_{x}/k_{B}$, Fermi energy $E_{F}/k_{B}$, temperature $T_{\mathrm{tof}}$, and tunneling time $\tau_J$.}
\label{tab:parameter_crossover_SM}
\begin{tabular}{lcccccccc}
	\hline\hline
	$s$ ($E_{R}$) & 12.75 & 15.30 & 17.00 & 25.50 & 34.00 & 42.50 & 51.00 & 59.50 \\
	\hline
	$\omega_{x}/2\pi$ (Hz) & 6315 & 6918 & 7293 & 8931 & 10313 & 11531 & 12631 & 13643 \\
	$\hbar\omega_{x}/k_{B}$ ($\mathrm{\mu K}$) & 0.303 & 0.332 & 0.350 & 0.429 & 0.495 & 0.553 & 0.606 & 0.655 \\
	$E_{F}/k_{B}$ ($\mathrm{\mu K}$) & 0.109 & 0.118 & 0.154 & 0.207 & 0.179 & 0.278 & 0.248 & 0.363 \\
	$T_{\mathrm{tof}}$ ($\mathrm{\mu K}$) & 0.159 & 0.185 & 0.227 & 0.286 & 0.282 & 0.415 & 0.295 & 0.529 \\
	$\hbar\omega_{x}/k_{B}T_{\mathrm{tof}}$ & 1.906 & 1.795 & 1.542 & 1.500 & 1.755 & 1.333 & 2.054 & 1.238 \\
	$\hbar\omega_{x}/E_{F}$ & 2.780 & 2.814 & 2.273 & 2.072 & 2.765 & 1.989 & 2.444 & 1.804 \\
	$\tau_J$($\mathrm{m s}$)  & 23.5 & 40.7 & 57.1 & 269 & $1.03\times10^3$ & $3.46\times10^3$ & $10.5\times10^3$ & $29.2\times10^3$ \\
	\hline\hline
\end{tabular}
\end{table*}

\begin{figure}[htbp]
\centering
\includegraphics[width=\columnwidth]{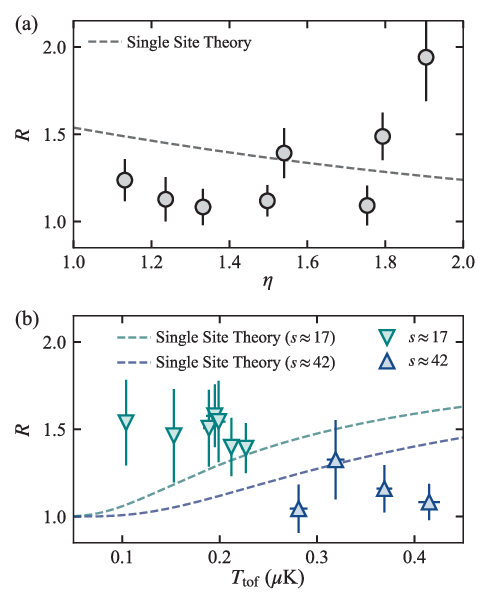}
\vspace{-15pt}
\caption{Comparison of the experimental branching ratio with the isolated single-site model. 
	(a) Branching ratio $R$ versus the thermal parameter $\eta = \hbar\omega_x / k_B T_\mathrm{tof}$. The dashed curve represents the theoretical prediction from Eq.~\ref{eq:single_site_formula}. 
	(b) Branching ratio $R$ as a function of $T_\mathrm{tof}$ for lattice depths $s \approx 17$ (downward triangles) and $s \approx 42$ (upward triangles). The dashed lines denote the corresponding single-site theoretical curves. In both panels, the experimental data show clear deviations from predictions. }
\label{p:SM_crossover}
\end{figure}

\section{Single-Site Thermal Model}
\label{sm:single_site}

To examine whether the observed evolution of the orbital branching ratio $R$ can be explained solely by local confinement within an isolated lattice site, we compare the experimental data with the prediction of a simple single-site thermal model. 

In an isolated harmonic trap, the parity selection rules of $p$-wave scattering allow the transverse $p_y$ and $p_z$ channels to be supported by all axial vibrational states, whereas the $p_x$ channel along the tight lattice axis strictly requires odd excitations ($n_x = 1, 3, 5, \dots$). Under a thermal Boltzmann distribution, the occupancy probability of the $n$-th axial state is given by $P_n = (1 - e^{-\eta}) e^{-n\eta}$, where $\eta = \hbar\omega_x/k_B T$. 
Summing over the discrete odd series yields the total population of all odd-parity states as $P_{\mathrm{odd}} = \sum_{m=0}^{\infty} P_{2m+1} = \frac{e^{-\eta}}{1+e^{-\eta}} = \frac{1}{e^{\eta} + 1}$. 
To ensure the model correctly retrieves the isotropic 3D limit ($R \to 2$) in the high-temperature limit ($\eta \to 0$), the effective weight of the $p_x$ channel is scaled as $2P_{\mathrm{odd}}$, while the unrestricted transverse channels each contribute a unit weight of $\sum_{n=0}^{\infty} P_n = 1$. Evaluating the branching ratio as $R(\eta) = (1 + 2P_{\mathrm{odd}})/1$ directly yields the analytical form:
\begin{equation}
R(\eta) = \frac{(1+e^{-\eta}) + 2e^{-\eta}}{1+e^{-\eta}} = 1 + \frac{2}{e^{\eta} + 1}.
\label{eq:single_site_formula}
\end{equation}

When $\eta\ll1$, thermal excitation allows all three orbital components ($p_x$, $p_y$, and $p_z$) to contribute nearly equally, yielding  $R\rightarrow2$. 
In contrast, when $\eta\gg1$, the axial excitation becomes energetically suppressed, leading to the freezing of the $p_x$ contribution and $R\rightarrow1$.

We first evaluate $\eta$ for the experimental data presented in Fig.~2(a) of the main text. 
As shown in Fig.~\ref{p:SM_crossover}(a), the measured values of $R$ do not collapse onto the theoretical curve predicted by Eq.~(\ref{eq:single_site_formula}). 
In particular, in the large-$\eta$ regime where the isolated-site model predicts a monotonic decrease of $R$ with increasing $\eta$, the experimental data do not follow the same trend.

This discrepancy becomes even more evident when examining the temperature dependence of $R$ at fixed lattice depths. 
Figure~\ref{p:SM_crossover}(b) shows the measured $R(T)$ at $s\approx17$ and $s\approx42$, together with the predictions of the single-site thermal model (dashed lines). 
In both cases, the experimentally observed evolution deviates systematically from the theoretical expectation.

\section{Tunneling-Induced Kinetic Anisotropy}
\label{sm:tunneling}

We interpret the evolution of the orbital branching ratio $R$ in terms of lattice-induced anisotropy of the open-channel relative kinetic energy. 
Under anisotropic confinement, the momentum distribution of the scattering states becomes direction dependent, leading to different projection weights onto the $p$-wave orbital channels~\cite{Kurlov2017}. 
The loss contribution of each orbital branch can therefore be estimated as
\begin{equation}
Q_3^{\ml} \propto \int d^3\mathbf{k}\,
n(\mathbf{k})\,
|Y_{1,\ml}(\hat{\mathbf{k}})|^2,
\label{eq:Q3_angle}
\end{equation}
where the spherical harmonics are defined with respect to the magnetic-field axis.

Using
$|Y_{1,0}|^2 \propto k_z^2/k^2$
and
$|Y_{1,\pm1}|^2 \propto (k_x^2+k_y^2)/k^2$,
the branching ratio can be approximated as
\begin{equation}
R \equiv \frac{Q_3^{1}}{Q_3^{0}}
\simeq
\frac{\langle k_x^2+k_y^2\rangle}
{\langle k_z^2\rangle}
=
\frac{
	\langle E_{\mathrm{kin},x}\rangle
	+
	\langle E_{\mathrm{kin},y}\rangle
}
{
	\langle E_{\mathrm{kin},z}\rangle
}.
\label{eq:R_Ekin}
\end{equation}

In the weak-confinement regime, the momentum distribution remains approximately isotropic, yielding
$\langle E_{\mathrm{kin},x}\rangle
=
\langle E_{\mathrm{kin},y}\rangle
=
\langle E_{\mathrm{kin},z}\rangle$
and therefore the 3D limit $R\simeq2$. 
As the lattice depth increases, motion along the lattice direction becomes progressively suppressed by the narrowing bandwidth of the lowest Bloch band, reducing the axial kinetic-energy contribution $\langle E_{\mathrm{kin},x}\rangle$. 
In the deep-lattice limit, this suppression drives the crossover toward the quasi-2D limit $R\rightarrow1$.

To characterize this crossover phenomenologically, we relate the suppression of axial motion to the tunneling amplitude $J(s)$ of the lowest Bloch band. 
For a sinusoidal optical lattice,
\begin{equation}
J(s)\propto s^{3/4}e^{-2\sqrt{s}},
\end{equation}
which decreases exponentially with lattice depth~\cite{Zwerger2003}. 
Motivated by this scaling behavior, we introduce the phenomenological form
\begin{equation}
R(s)
=
R_{\mathrm{2D}}
+
A_R e^{-\gamma\sqrt{s}},
\label{eq:R_s}
\end{equation}
where $R_{\mathrm{2D}}$ is the asymptotic quasi-2D limit, $\gamma$ characterizes the effective tunneling-induced decay rate, and $A_R$ is a free amplitude parameter. 
As shown in the main text, Eq.~(\ref{eq:R_s}) provides a good description of the experimental evolution of $R$.

~
\\

%-------------------------------------------------------------------------------------%

%\makeatletter
%\renewcommand{\@biblabel}[1]{[S#1]}
%\renewcommand{\@cite}[2]{[S#1\if@tempskipa ,S#2\fi]}
%\makeatother

\end{document}